\documentclass[letterpaper,aps,floatfix,twocolumn]{revtex4}
\usepackage{graphicx}
\usepackage{blindtext}
\usepackage{grffile}
\usepackage{epsfig}
\usepackage{amsmath}
\usepackage{amssymb}
\usepackage{hyperref}
\usepackage{alltt}
\usepackage{setspace}
\usepackage[english]{babel}
\usepackage{euscript}
\usepackage{tikz}
\usepackage[T1]{fontenc}
\usepackage[latin9]{inputenc}
\usepackage{yfonts}

\newcommand{\old}[1]{{}}

\begin{document}

\title{Kubo Combinatorics for Turbulence Scaling Laws}

\author{J. Petrillo}
\email{jarret.petrillo@stonybrook.edu}
\affiliation{Stony Brook University, Stony Brook, NY 11794, and GlimmAnalytics LLC, USA}
\author{J. Glimm}
\email{glimm@ams.sunysb.edu}
\affiliation{Stony Brook University, Stony Brook, NY 11794, and GlimmAnalytics LLC, USA}

\date{\today}

\begin{abstract}
We present an extension to Kolmogorov-Obukhov refined similarity hypotheses for universal fully-developed turbulence.
The extension is applied within Z.~She and E.~Leveque's multifractal model of inertial range scaling and its generalizations.
Our development has relevance to universal fully developed turbulence, a state we describe explicitly under the additional assumptions of the Kolmogorov-Obukhov similarity hypotheses in terms of the coupling between velocity fluctuations and averaged energy dissipation at all orders.
This description is unique and leads to a reparametrization of the She-Leveque model that preserves its original forecasts and is infinitely divisible.
\end{abstract}

\maketitle

\section{Motivation}

Velocity fluctuations and energy dissipation have played a large role in current investigations of turbulence.
The Kolmogorov-Obukhov refined similarity hypotheses \cite{Kol1962} link these variables in universal fully-developed turbulent flows by proposing a linear, first-order coupling.
Scaling models, which fit empirical data and succeed as models of chaotic turbulence, model regions more general than the linear law of the refined similarity hypotheses, but are still consistent with its model of energy cascades \cite{Obu1962, SheWay95}.
The multifractal extension envisions differentiated, singular, fractal-dimension regions that determine process decay in the surrounding fluid.

Our study is set within the multifractal formalism of Z.~She and E.~Leveque.
Reparametrizations to the She-Leveque model (SL) have been proposed based on an infinitely divisible law (Sec. \ref{sec:reparam}).
We also accept the Kolmogorov-Obukhov refined similarity hypotheses as a coupling between velocity fluctuations and energy dissipation that is linear in the velocity gradients and all fractal orders in the velocity differences.
We find a unique reparametrization of the SL parameters consistent with infinite divisibility and our interpretation.
Our result is compared to the model of S.~Chen and N.~Cao, who parameterize SL with the addition of observed turbulent data, and to the model of O.~Baratov.
The resulting parameters are consistent with the original SL values, shown in Fig. \ref{figch6}.

Our key contribution is the expression of the correlation between energy dissipation rates and velocity fluctuations as a nearly-disconnected graph.
We assume that velocity differences are correlated only through a region's average energy dissipation rate and viscosity, and that the linear coupling is homogeneous to all orders.
This analysis is based on a restriction to universal fully-developed turbulence, a state we label the \textbf{Kubo state}.
We show that this state is the one considered by M.~Parinello and T.~Arai in their application and extension of R.~Kubo's cumulant expansion method to infinite orders.

We rectify the Kolmogorov-Obukhov model of turbulent energy transfer with multifractal scaling.
We show that an interpretation of A.~Kolmogorov's refined similarity argument to all orders of the velocity differences leads to a unique reparameterization of the SL model that has virtually the same predictions and in addition is infinitely divisible.
The latter is a statistical requirement for a continuum of existent length scales that is not satisfied in the original SL model.

An extension of our results would model energy dissipation directly without the assumption of Kolmogorov-Obukhov refined similarity.
A theory without the Kolmogorov-Obukhov refined similarity law is not included in the present paper.
A complication arises in generalizing our result because in general the Kubo state does not satisfy independence of increments, a key requirement of a L\'{e}vy process (Secs. \ref{sec:intermittency}, \ref{sec:reparam}).
Our results can be interpreted as the leading order description for turbulence in the Kubo state, whereas next-to-leading order dynamics, modeled to an approximation in the original SL model, include non-L\'{e}vy couplings.

\section{Prior Studies}
\subsection{Intermittency}
\label{sec:intermittency}
Intermittency is a clustering of intense events that is evident at every order.
A first order cluster is the spatial distribution of a quantity of interest.
A second order cluster is a cluster of clusters.
In turbulence driven by the Navier-Stokes system, extreme events manifest as divergences in the velocity field \cite[p.~188]{Gal2002}.
Extreme values in the fluctuation of the velocity field are measurable by distributional changes in the two-point structure function, defined as
\begin{equation}
	\delta_{\vec{r}}\vec{u}(\vec{x},t)= \vec{u}(\vec{x}+\vec{r},t)-\vec{u}( \vec{x},t).
\end{equation}

The distribution of $\delta_{\vec{r}}(\vec{u})$ is dependent on location $\vec{x}$, the magnitude and direction of $\vec{r}$, and time $t$.
But, certain solutions of the Navier-Stokes equations, termed \textbf{universal fully-developed} states, exhibit an isometry and homogeneity that narrows these functional dependencies.
At fixed time in a universal fully-developed state, the structure function is dependent only on $r=||\vec{r}||$.

The moments
\begin{equation}
	\mu_n = E[( \delta_r \vec{u})^n]
\end{equation}
capture the spatial dependency of turbulent structures at different orders.
Intermittency of order $n$ is defined by the scaling in $r$ of $\mu_n$.

If 
\begin{equation}\label{eq:muzetan}
	\mu_n = C_n r^{\zeta_n}
\end{equation}
for $C_n$ a constant independent of $r$, then $\mu_n$ follows a scaling law in $r$ with exponent $\zeta_n$.

Equivalently,
\begin{equation}
	\zeta_n = \frac{\partial \log \mu_n}{\partial \log r}.
\end{equation}

A functional description of $\zeta_n$ would describe all orders of intermittency. 
In other words, it would describe how extreme events within Navier-Stokes turbulence are clustered in space.

For a summary of prior results, we refer to U.~Frisch \cite{Fri1980} for a detailed account of turbulence modeling prior to the She-Leveque \cite{SheLev94} theory and to his \cite{Fri1995} including that development.
We mention specifically the fundamental contributions of A.~Kolmogorov and O.~Obukhov \cite{Kol1941,Obu1962}, on fully-developed turbulence, contributions of \cite{RueTak71} on the transition to laminar flow, the roles of B.~Mandelbrot, E.~Novikov, A.~Sreenivasan, and many others in the development of these ideas, and the contribution of U.~Frisch in the development of the beta model.

The Kolmogorov-Obukhov refined similarity hypotheses underlie the K$62$ \cite{Kol1962} intermittency model, stated as
\begin{equation}
	\label{eq:k62}
	\text{K$62$: } \zeta_n = n/3 + \mu n(n-3)/18,
\end{equation}
for $\mu$, a scalar, termed the intermittency coefficient.
The space-average energy dissipation is defined for a ball of radius $r$ as:
\begin{equation}
	\epsilon_r(\vec{x}_0,t) = \frac{3}{4\pi r^3}\int_{B_r(\vec{x}_0)} \epsilon(\vec{y},t) d\vec{y}.
\end{equation}
The refined similarity hypotheses state that velocity fluctuations inside $B_r$ depend only on space-averaged dissipation over the same region, and that the space-averaged energy dissipation follows a log-normal process (See \cite[pp.~254-260]{Pop2000} for the relation to earlier theories).
Conceptually, it states that the expectation of $\delta_r \vec{u}$ is modelable as the tower property expectation of the conditional expectation of $\delta_r \vec{u}$ on $\epsilon_r$ \cite[p.~590]{MonYag75}:
\begin{equation}
	E[(\delta_r \vec{u})^n] = E[E[(\delta_r \vec{u})^n | \epsilon_r]],
\end{equation}
and that $\epsilon_r$ has a log-normal spatial distribution.
The log-normal distribution is importantly infinitely divisible \cite[p.~47]{Sat1990} (See Sec. \ref{sec:reparam}).

\subsection{She-Leveque Model}
\label{sec:she-leveque}

The She-Leveque model (SL) reasons constructively by an analogy between limiting dissipation regions and filamentary structures to show that \cite{SheLev94}
	\begin{equation}\label{eq:sl}
		\text{SL: } \zeta_n = \frac{n}{9} + 2 - 2\bigg(\frac{2}{3}\bigg)^{\frac{n}{3}}.
	\end{equation}

The quantities of interest are $\epsilon^{(n)}$, unobserved fractional scaling structures, that contribute to $\epsilon_r$, the local spatial-averaged energy dissipation in a ball of radius $r$.
The fractional structures $\epsilon^{(n)}$ are homogeneous in $r$ and are assumed to interact only with a limiting structure, $\epsilon^{(\infty)}$, and neighboring structures: $\epsilon^{(n+1)}$ and $\epsilon^{(n-1)}$.
The relationship is log-linear, and has the following functional form: \cite{SheLev94}
\begin{equation}
	\label{she-leveque-1}
\epsilon^{(n+1)} = A \epsilon^{(n)(\beta)} \epsilon^{(\infty)(1-\beta)},
\end{equation}
where $\beta$ is the production term and $A$ is a constant. 
In analogy with tubular structures, (assumptions that have been the object of much discussion) $\epsilon^{(\infty)}$ is assumed to scale in $r$ with an exponent of $-2/3$. 
Fractional scaling structures have a scaling exponent given by a difference equation.

If $\mu_n^{\epsilon}$ is the $n^{th}$ moment of $\epsilon_r$, defined as a spatial-average of energy dissipation in a ball of radius $r$, then the analogous exponent to $\zeta_n$ for $\epsilon_r$ is defined as:
\begin{equation}
	\label{eq:tau}
	\tau_n = \frac{\partial \log \mu_n^{\epsilon}}{\partial \log r}.
\end{equation}
SL assumes $\epsilon^{(n)}$ scales by $\tau_{n+1}-\tau_n$.  

The model's subsequent derivation comes from the substituition of these assumptions into the log-linear form.
The result is a second-order difference equation in $\tau_n$: 
\begin{equation}
	\tau_{n+2} - (1+ \beta) \tau_{n+1} + \beta \tau_n + \frac{2}{3} ( 1 - \beta) = 0.
\end{equation}
One solution is
\begin{equation}
	\tau_n = - \frac{2}{3} n + 2 - f(n)
\end{equation}
for $f(n)$, a function in $n$ that inherits log-linearity.
The difference equation has a unique solution under these assumptions:
\begin{equation}
	\label{eq:sl-tau}
	\tau_n = - \frac{2}{3} n + 2 - 2\bigg(\frac{2}{3}\bigg)^{n}.
\end{equation}

The SL model is an extension of the Kolmogorov-Obukhov theory, and relies on its translation from $\tau_n$ to $\zeta_n$, which can be stated as:
\begin{equation}
	\label{eq:kolognormal}
	\zeta_n = \frac{n}{3} + \tau_{\frac{n}{3}}.
\end{equation}

Finally, substitution of (\ref{eq:sl-tau}) into (\ref{eq:kolognormal}) yields the SL model, (\ref{eq:sl}).

We note that there are alternative models that have different than log-linear dependence, but that they are dynamically distinct from the SL model.  See \cite{Nel1995} and \cite{FriSul78}.

\subsection{Reparameterizations of She-Leveque}
\label{sec:reparam}

Different parameterizations of SL have been the subject of extensive research.
Using the notation of \cite{Bar1997, CheCao95}, the SL model (\ref{eq:sl}) can be rewritten with two free parameters, $x$, $C_0$, and a convenience term $\beta=(1-x/C_0)$:
\begin{equation}
	\zeta_n = \frac{n}{3}(1-x) + C_0( 1 - \beta^{n/3}). \label{eq:sl-general}
\end{equation}
The original SL parameterization is: $x=2/3$ and $C_0=2$.

There is a consensus that $x=1$ may be an improvement.
This criticism by E.~Novikov \cite{Nov1994} centers on a statistical argument.
A family of characteristic functions is infinitely differentiable if it is closed under integral roots \cite[p.~31]{Sat1990}.
In the context of length scales, it is a consistency requirement.  
Let $\{ \psi_{r}\}$  be a family of characteristic functions in $r$ associated with the two-point spatial distributions.  
Infinitely divisibility requires there exists an $r_q$ for every $r$ and integer $q$ so that 
\begin{equation}
	\psi_{r_q} = \sqrt[^q]{\psi_r}.
\end{equation}
Physically, it means that large length scales can be decomposed into smaller ones that together have the same distribution.
If $x=1$ then the property of infinite divisibility is fully satisfied.

Recall that an infinitely divisible distribution has a one-to-one identification with a L\'{e}vy process (L\'{e}vy-Khintchine formula) \cite[p.~13]{Ber1996} and that 
 a L\'{e}vy process has independent and identically distributed increments.
We note that $x\neq 1$ is an important piece in the original SL model because non-L\'{e}vy dynamics are an important part of turbulent dynamics away from the first mode.
Empirically, L\'{e}vy processes are used to model turbulent velocity fluctuations over all length scales \cite[Figs.~2-4]{NieBlaSch04}.
Their justification in modeling Navier-Stokes turbulence come from fractal Laplacian approximations to the governing system \cite{CheZho05}.
Turbulence generally exhibits a long-range dependence \cite{MorDel02} that is incompatible with independent increments.
Recall that the SL model relied on a restriction to universal fully-developed turbulence.
If added to this are the Kolmogorov-Obukhov similarity hypotheses, then indeed it was shown to restrict turbulence to a L\'{e}vy field (Sec. \ref{sec:intermittency}).
It is one of the major contributions of the present work to clarify L\'{e}vy solutions in turbulence (equivalent to the Kolmogorov-Obukhov theory of self-similarity) as the first mode in a more general solution.

Setting $x=1$, 
\begin{eqnarray}
	C_0 &=& \lim_{n \rightarrow \infty} \zeta_n, \label{key-relation}
\end{eqnarray}
which will become a key relation in our reparameterization.
Additionally, if 
the empirical intermittency correction
\begin{equation}\label{eq:sixthcorr}
\mu := 2 - \zeta_6
\end{equation}
as is common, then with $x=1$:
\begin{eqnarray}
	\mu &=&  2 - C_0( 1 - (1-1/C_0)^{2})\\
	&=&  C_0^{-1}. \label{taumu}
\end{eqnarray}
It follows that $C_0$ can be uniquely identified from an empirical estimate, $\hat{\mu}$.

$\hat{\mu}$ is an empirical quantity with estimated values ranging between $0.2$ and $0.45$ \cite[p.~2]{Bar1997}.

S.~Chen and N.~Cao \cite{CheCao95} estimate $\mu = \frac{2}{9}$, and set $C_0 = 4.5$ by (\ref{taumu}) and O.~Boratov \cite{Bar1997} reinterpreting \cite{Nel1995} estimate $\mu=0.41$, and set $C_0 = 2.43$.

\section{Kubo Combinatorics}
\label{sec:kubo}

Kubo combinatorics describe the nonlinear relationship between statistical cumulants and moments \cite{Kub1962}.
Their extension and application to infinite order systems was developed by M. Parinello and T. Arai \cite{ParAra74}.
From the original publications there was an implicit parallel between the graph-theoretic notion of connected and statistical dependence.

\subsection{Moments and Cumulants}

A characteristic function, denoted $\psi$, is the Fourier transform of a probability distribution (Sec. \ref{sec:reparam}).
The moments, $\mu_n$, are defined by a Taylor expansion of the characteristic function about zero as:
\begin{equation}
	\psi(X)|_{X=0} = \sum_{n=1}^{\infty} \mu_n \frac{X^n}{n!}.
\end{equation}
Similarly, the cumulants, $\kappa_n$, are defined as a Taylor expansion about zero of the log of the characteristic function:
\begin{equation}
	\log \psi(X)|_{X=0} = \sum_{n=1}^{\infty} \kappa_n \frac{X^n}{n!}.
\end{equation}
It is clear that $\kappa_n$ and $\mu_n$ are related.
It is a fact that each are related by a polynomial function that is heavily combinatorical.
Let $E_c[\cdot]$ be the first cumulant in parallel to the first moment, $E[\cdot]$.

For example, a normal distribution has moments:
\begin{equation}
	\mu_n = \sigma^n (-i\sqrt{2})^n U\Big(\frac{-n}{2},\frac{1}{2}, -\frac{1}{2}\Big(\frac{\sigma}{\mu}\Big)^2\Big),
\end{equation}
where $U$ is a confluent hypergeometric function, $\mu$ is the mean, and $\sigma^2$ is the variance.

In contrast, the cumulants of a normal distribution are $\kappa_1 = \mu$, $\kappa_2 = \sigma^2$, and $\kappa_p=0$ for $p>2$.

This stark example shows how natural it is to think in cumulants, although their use outside of distributional theory is not common.
In addition, the cumulant average of two distributions is zero if and only if the two distributions are independent (Kubo's theorem, Sec. \ref{sec:kubothm}).
From a practical standpoint, cumulants encode precisely the new information at every order of distributional structure.
A normal distribution is known to be parameterized by two parameters, and all cumulants of order three and higher are zero.

\subsection{Kubo's Theorem}
\label{sec:kubothm}

\begin{list}{}{\leftmargin=\parindent\rightmargin=0pt}
\item
	\textbf{Theorem} (Kubo).
Two groups of random variables, $\mathcal{G}_1$ and $\mathcal{G}_2$, are
independent if and only if every mixed cumulant average is zero. In particular,
a cumulant average is zero if its elements can be divided into two or more groups that are statistically independent.\\
{\it Proof.} See \cite{Kub1962}.
\end{list}

	Two random variables are {\bf statistically dependent} if their cumulant expectation is nonzero.
	More generally, two random variables are {\bf cumulant connected} if their cumulant expectation is nonzero.
	Cumulant connected implies statistically dependent.

\subsection{Parrinello-Arai Corrections}
\label{sec:infinite-connected}

A Parrinello-Arai correction is the multiplicative factor that accounts for an additional variable added in a system when the new variable adds one cumulant connection.
The technique was developed in \cite{ParAra74} as an extension to Kubo's theorem, and allowed the first nontrivial and accurate diagrammatic calculation for electron spin \cite[p.~98]{GelSin10}.

\begin{list}{}{\leftmargin=\parindent\rightmargin=0pt}
\item
	\textbf{Theorem} (Parrinello-Arai). Let $\mathcal{G}_m$ be an $m$-variable system such that all $X_j$ are statistically independent from $X_{k}$ for $k\neq1,j$.
	let $\mathcal{G}_{m+1}$ be a system with $m+1$ variables formed from $\mathcal{G}_m$ by the addition of $X_{m+1}$, a new variable that is cumulant connected to $X_1$ but statistically independent from $X_k$, for $k\neq1,m+1$.  Then 
	\begin{equation}
		E_c[ \mathcal{G}_{m+1}] = \Delta(X_{m+1}) E_c[ \mathcal{G}_m]
	\end{equation}
	where
	\begin{equation}
		\Delta(X_{m+1}) = \frac{E_c[X_{m+1}X_1]}{E[X_1]}.
	\end{equation}
	\textit{Proof.} \cite[p.~268]{ParAra74}
\end{list}

\begin{figure}\label{temp2}
\centering
\includegraphics[scale=0.2]{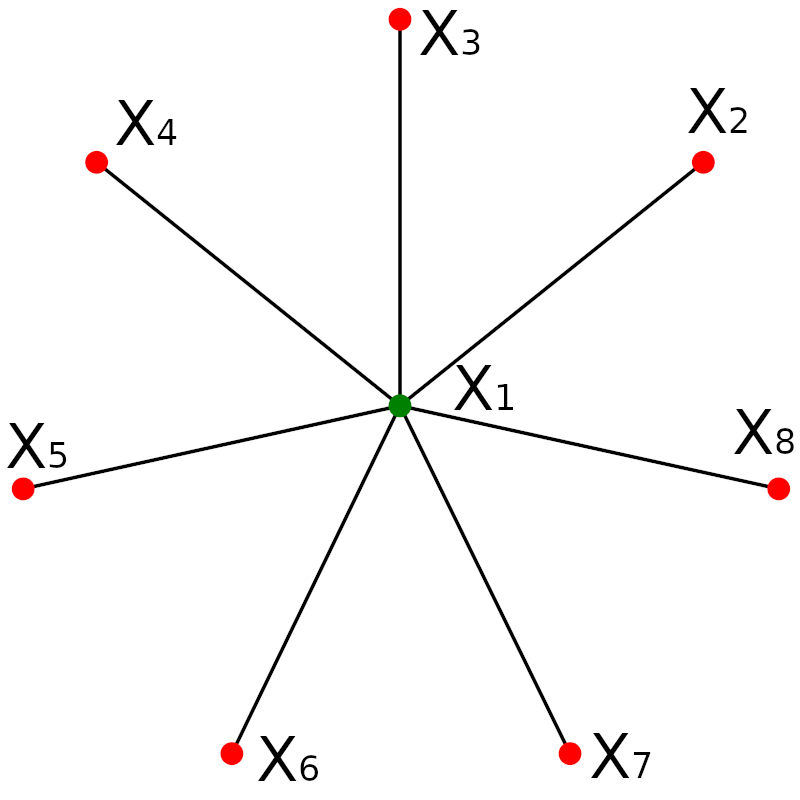}
	{\caption{{\bf Diagrammatic Representation of a Kubo State.} 
	In the diagrammatic representation connected nodes represent a cumulant connection.  The Kubo state has all variables cumulant connected to a central node, and otherwise has no cumulant connections.  The variable $X_1$ moderates all dependencies between system variables.}
	\label{figonestep}}
\end{figure}

One way to picture such a system is to imagine $X_1$ as the center and the $\{ X_2, \dots, X_{m+1}\}$ as terminal nodes of a free graph (Fig. \ref{figonestep}).
We term a system with this dependency structure a {\bf Kubo state}, and we will show how turbulent-relevant quantities have such a representation in universal fully-developed turbulence.

\label{temp2}

The Parinello-Arai theorem allows a cumulant to be written as a scalar product of a one-order-lower cumulant.
By repeating, the theorem allows full systems to be expressed as functions of the first, second, and
order-two mixed cumulant expectations.

Let us consider an infinite-dimensional Kubo state satisfying the conditions of the Parinello-Arai theorem: $\{ X_1, X_2, \dots \}$.
We write
\begin{equation}
\label{eq-chi-infty}
	X^{(\infty)} = \sum_{j=2}^{\infty} E_c[ X_j \dots X_1].
\end{equation}
This equation reads: $X^{(\infty)}$ is the limiting sum of all cumulant-connected subsystems, read configurations.

We posit that $X^{(\infty)}$ is the most important in the study of extrema because 
a system with connected components moves in tandem, reducing the effective degrees of freedom.
 The more degrees of freedom, the less the system deviates from mean behavior 
(recall that Chebycehv's inequality provides an upper bound on extreme events explicitly from a system's degrees of freedom \cite[p.~115]{Cho1994}).
For an infinite system, $X^{(\infty)}$ is the fully-connected state, and the one most likely to exhibit extreme behavior.

Let us write $X$ for $X \in \{X_2, X_3, \dots \}$ if the subsystem $\{X_2, X_3, \dots \}$ is homogeneous.
Homogeneous is used here to describe a system with statistically identical elements.
With the notation $X$ for the homogeneous element,
\begin{eqnarray}
	X^{(m+1)} &=& \sum_{j=2}^{m+1} E_c[ X_j X_{j-1} \dots X_1] \\
	&=& E_c[X X_1] \bigg( \sum_{j=0}^{m+1-2} (\Delta_1 X)^j \bigg)
\end{eqnarray}
where the last line is a geometric sum.  We have shown the following.
\begin{list}{}{\leftmargin=\parindent\rightmargin=0pt}
\item
	\textbf{Theorem} (Kubo Extension).
	Let $\mathcal{G}$ be an infinite-dimensional Kubo state satisfying the conditions of the Parinello-Arai theorem; let $X \in \mathcal{G}$ for all $X \neq X_1$ be statistically homogeneous. Then
\begin{equation}
	\label{infsum}
	X^{(\infty)} = E_c[X X_1] \bigg(1 - \frac{E_c[X X_1 ]}{E[X_1]}\bigg)^{-1}.
\end{equation}
\end{list}

\section{Application to Turbulent Scaling}
\label{sec:application-turbulence}

The Kubo extension theorem holds generally for Kubo states.
Kubo states can be understood to be a type of weakest-interaction approximation.
It is an infinite connected state that is one-disconnection away from being disconnected.

We proceed to substitute turbulent-relevant quantities and derive a reparameterization to the She-Leveque scaling law model.
We apply the analysis with $X_1$ equal to energy dissipation times viscosity and the other $X_j$ equal to velocity differences, averaged over some length scale.
The formulation is more general and not restricted to this choice.

Let us consider an infinite system $\mathcal{V}$ for fixed $\vec{x}$ in the turbulent domain $\mathbb{T}^3$, a fixed length scale $r$, and fixed time $t$:
\begin{equation}
	\mathcal{V} = \{ \nu \epsilon_r(\vec{x}), u_1(\vec{x}+\vec{r}_1)-u_1(x), u_1(\vec{x}+\vec{r}_2) - u_1(\vec{x},t), \dots \}
\end{equation}
where $||\vec{r}_j||=r$, $\epsilon_r(x,t)$ is the average turbulent energy dissipation rate in a ball of radius $r$ about $\vec{x}$,
and $\nu$ is the constant flow viscosity.
$u_1(\vec{x}+\vec{r}_j,t)-u_1(\vec{x},t) = \delta_{r_j} u_1$ are statistically homogeneous longitudinal velocity differences
that in our description are cumulant connected to $\epsilon_{r}$ but are otherwise pairwise independent variables.

This is a geometric image in agreement with, but an extension of the Kolmogorov-Obukhov refined similarity hypotheses \cite{Kol1962} used in the proof of the K$62$ refined scaling law.
The local energy dissipation rate (Sec. \ref{sec:intermittency}) determines the turbulent intensity, and correspondingly the statistical properties of velocity differences.
Inside the spherical region velocity fluctuations and local energy dissipation are coupled linearly.

Then
\begin{eqnarray}
	\mathcal{V}^{(\infty)} &=& E_c[\nu \epsilon_r \delta_r u_1] \bigg(1 - \frac{E_c[\nu \epsilon_r \delta_r u_1 ]}{E[\nu \epsilon_r]}\bigg)^{-1}.
\end{eqnarray}
This equation describes the Kubo connected state of $\mathcal{V}$.
It contains the first moment of the velocity difference and a covariance term
\begin{equation}
	E_c[\nu \epsilon_r \delta_r u_1] = E[\nu \epsilon_r \delta_r u_1] - E[\nu \epsilon_r]E[\delta_r u_1].
\end{equation}

We seek an approximation to the scaling law for $\mathcal{V}^{(\infty)}$
by taking the logarithm and differentiating by $\log r$.

\begin{eqnarray}
	\log \mathcal{V}^{(\infty)} &=& \log E_c[\nu \epsilon_r \delta_r u_i] - \log \Big( \big<\nu \epsilon_r \big> - E_c[\nu \epsilon_r \delta_r u_i]\Big) \nonumber \\
	&& +\log \big<\nu \epsilon_r \big>\\
	\frac{ \partial \log \mathcal{V}^{(\infty)}}{ \partial \log r} &=& \frac{\partial \log E_c[\nu \epsilon_r \delta_r u_1]}{\partial \log r} - \\ \nonumber
&& \frac{\partial \log ( \big< \delta_r u_1 \big> - E_c[\nu \epsilon_r \delta_r u_1])}{\partial \log r} \nonumber \\ 
	&& + \frac{\partial \log \big< \nu \epsilon_r \big> }{\partial \log r}. \label{best}
\end{eqnarray}
We investigate each of the three terms in (\ref{best}) separately.

Within the She-Leveque framework the third term
$$\frac{ \partial \log \big< \nu \epsilon_r \big>}{\partial \log r} := \tau_1$$
is the scaling exponent of the first moment of average energy dissipation (see (\ref{eq:tau})).
This holds by $\nu$ a constant.
$\tau_1=0$ for stationary flows, which implies that the third term is zero.

The first term has a scaling exponent by dimensional considerations.
Recall that $\delta_r u_1$ has units $\Big(\frac{\text{length}}{\text{time}}\Big)$ and $\nu \epsilon_r$ has units $\Big(\frac{\text{length}}{\text{time}}\Big)^4$ \cite[p.~249]{Kam2007}.
It follows that $\nu \epsilon_r \delta_r u_1$ has units
\begin{equation}
\Big(\frac{\text{length}}{\text{time}}\Big)^5,
\end{equation}
It follows that
\begin{equation}
	\frac{ \partial \log E_c[\nu \epsilon_r \delta_r u_1]}{\partial \log r} = 5.
\end{equation}

\begin{figure}\label{temp}
\centering
\includegraphics[scale=0.65]{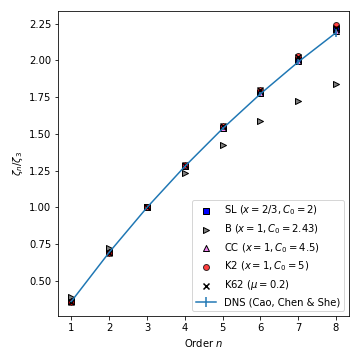}
	{\caption{{\bf Scaling Model Comparison.} 
	References: (SL) \cite{SheLev94}, (CC) \cite{CheCao95}, (B) \cite{Bar1997}, (K$62$) \cite{Kol1962}, (K2) [see text], and (DNS) \cite{CaoCheShe96}.  Direct numerical simulations (DNS) provided for empirical comparison.  With the exception of (B), all models align closely to this empirical sample.}\label{figch6}}
\end{figure}\label{temp}
\begin{table}
\begin{center}
\begin{tabular}{|l|l|l|l|}
	\multicolumn{1}{c}{\bfseries Order } & \multicolumn{1}{c}{\bfseries K62 ($\mu=0.2$)} & \multicolumn{1}{c}{\bfseries SL} & \multicolumn{1}{c}{ \bfseries K2 ($\mu=0.2$)}  \\ \hline
	6 & 1.80 & 1.77 & 1.80\\ \hline
	7 & 2.02 & 2.00 &  2.03 \\ \hline
	8 & 2.22 & 2.21 &  2.24 \\ \hline
	9 & 2.40 & 2.41 &  2.44 \\ \hline
\end{tabular}
\end{center}
	{\caption{{\bf Scaling Model Comparison}. With $\mu=0.2$, K$62$ and K$2$ have identical scalings for order $n=6$.  For order $n=8$, K$62$ and SL are nearly indistinguishable.} \label{table}}
\end{table}

\begin{figure}\label{temp2}
\centering
	\includegraphics[scale=0.65]{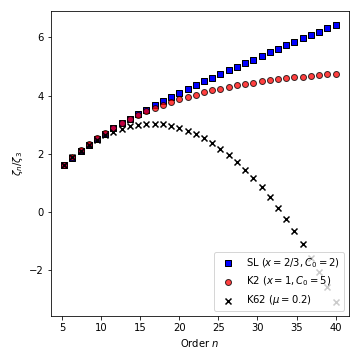}
	{\caption{{\bf Comparison Scaling Model Saturation Limit.} 
	References: (SL) \cite{SheLev94}, (K62) \cite{Kol1962}, and (K2) [see text].  The three models are differentiated by higher-order predictions.}\label{figch6_out}}
\end{figure}\label{temp3}

We now turn to the second term.
\begin{equation}
-\frac{\partial \log ( \big< \delta_r u_1 \big> - E_c[\nu \epsilon_r \delta_r u_1])}{\partial \log r} \nonumber
\end{equation}
Recall that power law distributions do not form an additively closed group.
Recall equation (\ref{eq:muzetan}).
In universal fully-developed turbulence (applicable to the SL model) distributional moments are power laws in the length scale.
The second term in (\ref{best}) is not a power law.
In the SL framework (see section \ref{sec:she-leveque}), certain logarithms in the limit are assumed to scale with a single fractal order.
This can be seen in (\ref{she-leveque-1}) by the fixed limit scaling behavior of $\epsilon^{(\infty)}$.
Therefore, the second term is anomalous scaling from the perspective of the SL theory.
Its existence represents the fact that the refined similarity scaling law is necessarily a restriction to the first mode of universal fully-developed turbulence.

We note that by velocity distributions convergent, the sign of the second term is negative.
This means that anomalous scalings, which are present, are single-directional deviations from the first mode described by our model.
Therefore, our model suggests that:

\begin{enumerate}
	\item Within the She-Leveque theory, the parameterization consistent with the Kolmogorov-Obukhov theory has $x=1$ and
		\begin{equation}
			C_0 = \lim_{n \rightarrow \infty} \zeta_n = 5.
		\end{equation}
	\item Universal fully-developed scaling in the Navier-Stokes system is bounded from above by 
		\begin{equation}
			\zeta_n = 5(1-(4/5)^{n/3}).
		\end{equation}
\end{enumerate}

We label our $x=1,C_0=5$ model as K2.
Fig. \ref{figch6} compares it to She-Leveque (SL), Chen-Cao (CC), Baratov (B), and direct numerical simulation \cite[table~1]{CaoCheShe96} (DNS).
We find that the K$62$, She-Leveque, and K2 models are nearly indistinguishable (Table \ref{table}) up to order $n = 12$ (Fig. \ref{figch6_out}).
Beyond order $n=12$, each model is differentiated by a positive infinite saturation limit (SL), a finite saturation limit (K2), and a negative infinite limit (K62).
The Chen-Cao, She-Leveque, K2, and K62 models align with empirical estimates.

\section{Conclusion}
The Kubo K2 theory extends Kolmogorov's inertial range coupling in universal fully-developed turbulence to the velocity saturation limit observed in S.~Chen and N.~Cao's interpretation of the SL model.
With the K2 theory, we have a reparameterization of She-Leveque (SL) with $x=1$, $C_0=5$ with an infinitely divisible law that preserves its efficacy in explaining empirical data.
The Kubo extension theorem is essential and holds in more generality than our application.
We have used it to extend a two-point structure correlation to correlations between all velocity fluctuations and averaged energy dissipation rate at a single length scale.
In this light, the extension theorem can be interpreted as our refinement to the Kolmogorov-Obukhov refined similarity hypotheses.

The She-Leveque, Chen-Cao, Baratov, and K2 models are set within the She-Leveque formalism and the same phenomenology applies.
These are models for universal fully-developed turbulence, a state that exhibits power-law scaling in its distributional moments, a homogeneity of velocity fluctuations, and a rotational symmetry that allows $|\vec{r}|=r$.
The Chen-Cao, Baratov and K2 models set $x=1$ and are therefore restrictions to an infinitely divisible substate.
The substate is an approximation by the first mode, and is consistent with A.~Kolmogorov's K$62$ model.

When we denoted K2 with the number 2 we were referencing the fact that the Kolmogorov-Obukhov refined similarity hypotheses assume a log-normal distribution in the second moment of $\epsilon_r$.
We frame this as a possibily ad-hoc preference.
We envision a new class of models within the She-Leveque and refined similarity theories i.e. K$N$,
where the new models are rederived with a preferential dissipation rate in order $N$.
We draw on emprical evidence from a study by J.~Glimm, R.~Kaufman, and A.~Hsu \cite[p.~11,Fig.~12]{GliKauHsu20} that finds a cubic, read nonlinear, dependence between the $n^{th}$ order scaling of local energy dissipation and $\tau_2$ for orders $n \leq 15$ to conclude that this proposed class of models will be nontrivial extensions.
Our working hypothesis, of which further computation and empirical investigation may confirm, is that K6 will have strong experimental efficacy.

It is interesting to speculate what type of experimental flow is universally fully-developed in the sense we propose.
Our reparametrization yields by (\ref{eq:sixthcorr}) an intermittency correction equal to $\mu = 0.20$: a value equal to empirical estimates from turbulent flows in atmospheric surface layers \cite{AntEtAl81, ChaAnt84}, turbulent planes and circular jets \cite{AntEtAl82}.

Returning to the derivation of K2 and (\ref{best}), recall that the second term of the limiting expansion is anomalous scaling.
This term models flow and time dependent relaxations to an infiniitely divisible first mode within universal fully-developed turbulence.
The Chen-Cao and Baratov models are recovered when the anomalous scaling term is set to $0.50$ and $2.57$, respectively.

\bibliographystyle{siam}
\bibliography{refs}

\end{document}